\documentclass[aps,11pt,prl,notitlepage,tightenlines,nofootinbib,twocolumn,superscriptaddress]{revtex4-2}
\usepackage{amsmath}
\usepackage{amssymb,amsthm}
\usepackage{mathtools}
\usepackage{mathrsfs}
\usepackage{relsize}
\usepackage{bm}

\usepackage{epsfig,graphics,graphicx,color,xcolor}
\usepackage{soul} 
\usepackage{cancel} 
\usepackage{slashed}	
\usepackage{verbatim}
\usepackage{ulem}
\usepackage{subfigure}
\usepackage{array}
\usepackage{ragged2e}
\usepackage{lineno}
\usepackage{natbib}
\usepackage{hyperref}
\hypersetup{colorlinks=true,linkcolor=red,anchorcolor=red,citecolor=orange, filecolor=brown,urlcolor=red,bookmarksnumbered=true,
pdfview=FitB
}
\graphicspath{ {image/} }
\usepackage{enumitem}

\colorlet{darkgreen}{green!50!black}
\colorlet{brightyellow}{yellow!75!red}
\colorlet{orange}{red!50!yellow}
\colorlet{darkgray}{gray!50!black}

\def\dd{{\mathrm{d}}}

\newcommand{\half}[1][1] {\mathsmaller{\frac{#1}{2}}}

\makeatletter
\newcommand*{\transpose}{%
  {\mathpalette\@transpose{}}%
}
\newcommand*{\@transpose}[2]{%
  \raisebox{\depth}{$\m@th#1\intercal$}%
}
\makeatother 



\begin{document}

\title{Chiral sum rule on the light front and the 3D image of the pion}

\author{Yang~Li}
\email{leeyoung1987@ustc.edu.cn}
\affiliation{Department of Modern Physics, University of Science and Technology of China, Hefei 230026, China}

\author{Pieter~Maris}
\affiliation{Department of Physics and Astronomy, Iowa State University, Ames, IA 50011}

\author{James~P.~Vary}
\affiliation{Department of Physics and Astronomy, Iowa State University, Ames, IA 50011}

\date{\today}

\begin{abstract}
The lightest meson, the pion, features two faces -- one is the elementary Goldstone boson of QCD and the other is the structured bound state of quarks and gluons. To accommodate both in a single light-front wave function in the valence space, we obtain a sum rule by analyzing the conserved axial-vector current  and the general structures of the wave functions. Using an analytic model motivated by holography, we show this sum rule is consistent with requirements of chiral symmetry breaking in AdS/QCD. Within this model, we find a remarkable feature of the pion, namely that the density is mostly uniform inside its radius; furthermore, we obtain good agreement with the experimental pion form factor at spacelike momenta.
\end{abstract}
\maketitle

\section{Introduction \label{sec1}} 
It is well known that the pion, the lightest meson, serves a dual role in Nature.  On the one hand, it is the Goldstone boson of the spontaneously broken chiral symmetry \cite{Nambu:1961tp, Goldstone:1962es}, as described in detail by the chiral effective field theory \cite{Weinberg:1978kz}. 
On the other hand, it is a composite particle -- like all other hadrons -- quarks and gluons bound together by the strong force. 
Perturbative QCD analysis of hard exclusive processes firmly establishes dominant contributions in the pion to the valence (quark and antiquark) Fock sector
\cite{Lepage:1980fj}. 

How does the structure of the pion accommodate these seemingly opposing perspectives?  Since confinement and chiral symmetry breaking are the two most fundamental 
features of QCD in the non-perturbative regime, the understanding of the strong interaction entails an answer to this question. 

Quark confinement from lattice gauge theory (LGT) is based on simulation of pure gauge theory with infinitely heavy quarks \cite{Wilson:1974sk}. For chiral fermions, lattice simulations are computationally formidable, and  no simple picture has emerged for the inner structure of the pion as a Goldstone boson \cite{Aoki:2021kgd}. 
Careful analysis of the axial-vector Ward-Takahashi identity (AVWTI) in Dyson-Schwinger equations (DSEs) shows that the dominant (pseudoscalar) component of the pion Bethe-Salpeter amplitude is equal to the dynamically generated scalar quark self-energy \cite{Maris:1997hd}; in addition, the AVWTI provides rigorous constraints on the other covariant structures of the pion as well. Indeed, models based on this analysis are very successful in describing light meson properties \cite{Maris:2003vk}.
However, the access to the partonic structure of the pion is still indirect, since the DSEs (as well as LGT) are defined in the Euclidean spacetime. 
An interesting perspective is provided by AdS/QCD \cite{Erlich:2005qh}. In this model, the two faces of the pion, confinement and chiral symmetry breaking, are encoded in the large-$z$ and small-$z$ behaviors in terms of the fifth dimension $z$, respectively, which is thought of as the (inverse) renormalization scale \cite{Gherghetta:2009ac}. 

Our modern understanding of the hadron structure is based on the parton picture, which corresponds to how the pion is ``seen'' in high-energy scattering experiments \cite{Brodsky:1997de, E791:2000xcx, Ashery:2003cd}.
In particular, the partonic structure of the pion is unambiguously encoded in its wave functions defined at fixed light-front time $x^+ = t + z/c$. These light-front wave functions (LFWFs) are amplitude-probabilistic, Minkowskian and frame independent.
There have been major efforts to extract these quantities from Euclidean quantum field theories, e.g. LGT and DSEs  \cite{Chang:2013pq, Shi:2018zqd, Ji:2020ect, Roberts:2021nhw, dePaula:2022pcb}.
On the other hand, in principle, the LFWFs can be directly obtained from diagonalizing the QCD Hamiltonian on a null plane $\omega \cdot x = 0$ with $\omega_\mu \omega^\mu = 0$ \cite{Brodsky:1997de}. This approach has recently been revived to solve quantum field theories in lower dimensions \cite{Anand:2020gnn}. 

Quark confinement entails confining interactions between partons at large separations in the pion. Then the question becomes: 
where is the chiral symmetry breaking facet in the pion wave functions? 
Inspired by the exact relations from the covariant analysis of the AVWTI in DSEs, we show there is also a relation, a chiral sum rule, on the pion LFWFs.  
The sum rule developed below reads,
\begin{equation}\label{eqn:chiralSR}
\int \frac{\dd x}{\sqrt{x^3(1-x)^3}} \int \frac{\dd^2k_\perp}{(2\pi)^3} k_\perp^2 \psi_{\pi}(x, \vec k_\perp) = 0,
\end{equation}
in the chiral limit ($m_q \to 0$), where $\psi_{\pi}(x, \vec k_\perp)$ is the pion spin-singlet valence sector LFWF which depends on the relative momenta $(x, \vec k_\perp)$ between the quark and the antiquark (see definition below).
We further show, in the context of light-front holography \cite{Brodsky:2014yha}, this sum rule is equivalent to the chiral condensate in AdS/QCD. 

Note that here we neglected the difference in the bare (current) $u,d$ quark masses and consider the unflavored case with $m_q = \frac{1}{2}(m_u + m_d)$. We defer the fully consistent treatment of the UV including the  renormalization to future work and focus on the non-perturbative part of the physics, confinement and chiral symmetry breaking. Therefore, we have suppressed the explicit dependence on the renormalization scheme and scale.

\section{Formalism \label{sec 2}}

The key point for obtaining this sum rule (\ref{eqn:chiralSR}) is the covariant decomposition of the pion LFWFs. In Fock space, the state vector of a pseudoscalar $P$ can be decomposed as, 
\begin{multline}
|P(p)\rangle =\sum_{s,\bar s}\int\frac{\dd x}{2x(1-x)}\int \frac{\dd^2k_\perp}{(2\pi)^3} \psi_{s\bar s/P}(x, \vec k_\perp) \\ 
\times  \frac{1}{\sqrt{N_C}}   \sum_i b^\dagger_{si}(p_1)d^\dagger_{\bar si}(p_2)|0\rangle 
+ \cdots
\end{multline}
where the ellipsis represents the higher Fock sector wave functions.  Here $N_C=3$ is the number of colors. $p_1$ and $p_2$ are the quark and antiquark momenta, respectively. 
We adopt the standard light-front coordinates, where the components of a 4-vector $v$ are written as $v^\pm = v^0 \pm v^3$ and $\vec v_\perp = (v^1, v^2)$, corresponding to $\omega^-=2$, $\omega^+ = \omega_\perp = 0$ \cite{Brodsky:1997de}. 
The longitudinal momentum fraction 
of the quark (antiquark) is $x = p_1^+/p^+$ ($1-x=p_2^+/p^+$). The relative transverse momentum $\vec k_\perp = \vec p_{1\perp} - x\vec p_\perp$. Momentum conservation implies, $p^+ = p^+_1 + p_2^+$ and $\vec p_\perp = \vec p_{1\perp} + \vec p_{2\perp}$. 
The most general form of the pion valence ($q\bar q$) LFWF $\psi_{s\bar s/P}(x, \vec k_\perp)$ that satisfies all kinematical symmetries on the light front is given by the covariant light-front dynamics (CLFD) \cite{Carbonell:1998rj}, 
\begin{multline}\label{eqn:CLFD}
\psi_{s\bar s/P}(x, \vec k_\perp) = 
\bar u_{s}(p_1)\Big[\gamma_5 \phi_1(x, k_\perp) \\
 +
 \frac{\gamma_5\slashed\omega}{\omega\cdot p} \hat f_\chi \phi_2(x, k_\perp)\Big] v_{\bar s}(p_2).
\end{multline}
It depends on the orientation of the quantization surface (the light front), as indicated by the null vector $\omega$. Note that the null vector $\omega$ is only defined up to a scaling factor. Therefore, the conformal invariant structure $\slashed\omega/(\omega\cdot p)$ is mandatory. As a result, we need to introduce a constant $\hat f_\chi$ with mass dimension to balance the dimension. In CLFD, this constant is usually taken as the quark mass $m_q$ (or the pion mass $M_P$) which vanishes in the chiral limit \cite{Dziembowski:1986dr, Choi:1997iq, Carbonell:1998rj, Leitner:2010nx}. We instead take a constant non-vanishing in the chiral limit. Later, it will be shown that a convenient choice is the pion decay constant. 
Eq.~(\ref{eqn:CLFD}) can be written explicitly as, 
\begin{align}
\psi_{\uparrow\uparrow/P}(x, \vec k_\perp) =\,& \psi^*_{\downarrow\downarrow/P}(x, \vec k_\perp)  \label{eqn:triplet} \\
& = - \frac{k_\perp e^{-i\arg\vec k_\perp}}{\sqrt{x(1-x)}}\phi_1(x, k_\perp); \nonumber\\
\psi_{\uparrow\downarrow - \downarrow\uparrow/P}(x, \vec k_\perp) =\,& \frac{\sqrt{2}m_q}{\sqrt{x(1-x)}} \phi_1(x, k_\perp)  \label{eqn:singlet} \\
& - \hat f_\chi \sqrt{8x(1-x)}\phi_2(x, k_\perp); \nonumber
\end{align}
where $\psi_{\uparrow\downarrow - \downarrow\uparrow/P} = \big[\psi_{\uparrow\downarrow/P} - \psi_{\downarrow\uparrow/P}\big]/\sqrt{2}$.

\section{Chiral sum rule \label{sec 3}} 

The sum rule (\ref{eqn:chiralSR}) follows directly from examining the partially conserved axial-vector current (PCAC) \cite{Gell-Mann:1960mvl}, 
\begin{equation}\label{eqn:PCAC}
\partial_\mu J^\mu_5 = 2i m_q\overline q\gamma_5 q.
\end{equation}
Here, $J^\mu_5 = \overline q \gamma^\mu \gamma_5 q$ is the axial-vector current, and $J_5 = \bar q i\gamma_5 q$ is known as the pseudoscalar current. We have neglected the chiral anomaly term, which is irrelevant to our discussion here. Consider the local vacuum-to-pseudoscalar matrix elements of the operators,
\begin{align}
& \langle 0 | J^\mu_5(x) | P(p) \rangle = e^{-ip\cdot x} i p^\mu f_P, \\
& \langle 0 | J_5(x) | P(p) \rangle = e^{-ip\cdot x} g_P.
\end{align}
The constant $f_P$ is known as the decay constant. Applying the PCAC (\ref{eqn:PCAC}), we obtain the Gell-Mann-Oakes-Renner (GMOR) relation \cite{Gell-Mann:1968hlm}:
\begin{equation}\label{eqn:GMOR}
M_P^2f_P = 2m_q g_P, 
\end{equation}
where $M_P$ is the mass of the pseudoscalar, $M_P^2 = p^2$. 
In the LFWF representation, these local matrix elements are, 
\begin{align}
\langle 0 | J^\mu_5(0) |P(p)\rangle =\,& \sqrt{N_C}\sum_{s,\bar s}\int \frac{\dd x}{2x(1-x)}\int \frac{\dd^2k_\perp}{(2\pi)^3} \nonumber \\
\times \, & \psi_{s\bar s/P}(x, \vec k_\perp) \bar v_{\bar s}(p_2) \gamma^\mu\gamma_5 u_{s}(p_1), \\
\langle 0 | J_5(0)  |P(p)\rangle  =\,& \sqrt{N_C} \sum_{s,\bar s}\int\frac{\dd x}{2x(1-x)}\int \frac{\dd^2k_\perp}{(2\pi)^3} \nonumber \\
\times \, & \psi_{s\bar s/P}(x, \vec k_\perp) \bar v_{\bar s}(p_2)i\gamma_5 u_s(p_1).
\end{align}
Note that these expressions are exact, viz. the vacuum-to-pion matrix elements only depend on the valence wave function. Applying the PCAC (\ref{eqn:PCAC}) and the covariant decomposition of the LFWF (\ref{eqn:CLFD}), we obtain, 
\begin{multline}\label{eqn:PCAC_LFWF}
 \int \frac{\dd x}{2x(1-x)} \int \frac{\dd^2k_\perp}{(2\pi)^3}  \Big( \frac{k_\perp^2+m_q^2}{x(1-x)} - M_P^2\Big)  \\
  \times  \Big( 2m_q\phi_1(x, k_\perp) - 4x(1-x)\hat f_\chi \phi_2(x, k_\perp)\Big) = 0. 
\end{multline}
%
In the chiral limit $m_q \to 0$, this expression reduces to,
\begin{equation}\label{eqn:PCAC_LFWF_0}
\int \frac{\dd x}{2x(1-x)} \int \frac{\dd^2k_\perp}{(2\pi)^3} k_\perp^2 \phi_2^{(0)}(x, k_\perp) = 0,
\end{equation}
due to the GMOR relation (\ref{eqn:GMOR}), where we have used the superscript (0) to indicate quantities in the chiral limit. From Eq.~(\ref{eqn:singlet}), 
the leading-twist ($L_z=0$) wave function $\psi_\pi \equiv \psi^{(0)}_{\uparrow\downarrow-\downarrow\uparrow} \propto \sqrt{x(1-x)}\phi_2^{(0)}$ in the chiral limit. Hence, we obtain the chiral sum rule  Eq.~(\ref{eqn:chiralSR}). 

In the above derivation, there is an explicit assumption that $\hat f_\chi$ does not vanish in the chiral limit. 
If, instead, we take $\hat f_\chi \to 0$ as usually done in CLFD (cf. Ref.~{\cite{Jaus:1999zv}}), Eq.~(\ref{eqn:PCAC_LFWF}) will be automatically satisfied in the chiral limit -- for arbitrary pion LFWF. However, the pion decay constant (\ref{eqn:fpi}) also vanishes, and the pion mass is not required to be zero (\ref{eqn:GMOR}). Indeed, this is the case when the chiral symmetry is not spontaneously broken, and the pion is not a Goldstone boson regardless of its mass. 
 
In the vicinity of the chiral limit, we can expand the wave functions and related quantities in terms of the quark mass $m_q$, 
\begin{equation}
\phi_{1,2} = \phi_{1,2}^{(0)} + m_q \phi_{1,2}^{(1)} + \cdots.
\end{equation} 
Substituting this expansion into Eq.~(\ref{eqn:PCAC_LFWF}) and taking terms up to  $\mathrm{O}(m_q)$, we obtain, 
\begin{multline}\label{eqn:O1}
-2m_q   \int \frac{\dd x}{2x(1-x)} \int \frac{\dd^2k_\perp}{(2\pi)^3} \frac{2k_\perp^2}{x(1-x)} \phi_{1}^{(0)}(x, k_\perp) \\
 = M_P^2  4\hat f_\chi  \int \dd x \int \frac{\dd^2k_\perp}{(2\pi)^3}   \phi_{2}^{(0)}(x, k_\perp).
\end{multline}
The r.h.s. is just the wave function representation of the pseudoscalar decay constant in the chiral limit \cite{Lepage:1980fj}: 
\begin{equation}\label{eqn:fpi}
f_P^{(0)} = 4i\hat f_\chi \sqrt{N_C} \int \dd x \int \frac{\dd^2k_\perp}{(2\pi)^3}   \phi_2^{(0)}(x, k_\perp), 
\end{equation}
where the decay constant is extracted from the local matrix element $\langle 0 | J^+_5 | P(p)\rangle \equiv i p^+ f_P$. 
Similarly, the l.h.s. is the wave function representation of the pseudoscalar amplitude in the chiral limit \cite{Lepage:1980fj}:
\begin{multline}
g_P^{(0)} 
= -2i\sqrt{N_C} \int\frac{\dd x}{2x(1-x)}\int \frac{\dd^2k_\perp}{(2\pi)^3} \\
\times \frac{k_\perp^2}{x(1-x)}\phi_1^{(0)}(x, k_\perp).
\end{multline}
Therefore, Eq.~(\ref{eqn:O1}) is consistent with the GMOR relation (\ref{eqn:GMOR}), 
\begin{equation}\label{eqn:GMOR_O1}
M_P^2 f_P^{(0)} = 2m_q g_P^{(0)} + \mathrm O(m_q^2).
\end{equation}
Note that the above relation also implies that the decay constants of the excited pions vanish in the chiral limit \cite{Maris:1997hd}.

\section{Light-front holography \label{sec 4}} 

The effect of the chiral sum rule can be better pictured in a simple analytic model.  
In the chiral limit, the light-front kinetic energy depends only on a transverse vector $\vec\kappa_\perp = \vec k_\perp/\sqrt{x(1-x)}$. 
In light-front holography (LFH), a semi-classical approach to QCD \cite{Brodsky:2014yha}, it is assumed that the dynamics in the chiral limit 
is solely dictated by the transverse direction, and the wave function is separable \cite{Li:2021jqb, deTeramond:2021yyi, Li:2022izo, Lyubovitskij:2022rod, Ahmady:2021yzh, Ahmady:2021lsh}. Taking this ansatz, we can rewrite $\phi_2$ for 
a pseudoscalar $P$ as,
\begin{equation}\label{eqn:phi2_LFH}
\phi_{2,P}(x, k_\perp) = \frac{\pi}{2\sqrt{N_C}}\frac{\phi_P(x)}{x(1-x)f_\pi} \varphi_P(\kappa_\perp).
\end{equation} 
Here, $\phi_P(x)\propto \sqrt{x(1-x)}$ is the light-cone distribution amplitude with $\int \dd x \, \phi_P(x) = f_P$. $\varphi_P(\kappa_\perp)$ is the transverse function, whose Fourier
transform, 
\begin{equation}
\widetilde\varphi_P(\zeta_\perp) = \int \frac{\dd^2\kappa_\perp}{(2\pi)^2} e^{i\vec \kappa_\perp\cdot\vec\zeta_\perp}\varphi_P(\kappa_\perp),
\end{equation}
where $\vec \zeta_\perp = \sqrt{x(1-x)}\vec r_\perp$. %
The chiral sum rule (\ref{eqn:PCAC_LFWF_0}) implies, 
\begin{equation}\label{eqn:LFH_CSR}
 f_P \nabla^2_\perp \widetilde\varphi_P(\zeta_\perp=0) = 0.
\end{equation}
N.B. this relation is applicable to both the ground state, i.e. the pion, and the excited states, i.e. excited pions. For the pion, $\nabla^2_\perp \widetilde\varphi_P(\zeta_\perp=0) = 0$ is required. 
For the excited pions, the decay constants vanish in the chiral limit $f_P = 0$, as required by the GMOR relation (\ref{eqn:GMOR}) \cite{Maris:1997hd}.  

In LFH, the transverse wave function satisfies the semi-classical light-front Schrödinger wave equation,
\begin{equation}\label{eqn:LFSWE}
\Big[-\nabla^2_\perp + U(\zeta_\perp) \Big]\widetilde\varphi_P(\zeta_\perp) = M^2_P\widetilde\varphi_P(\zeta_\perp).
\end{equation}
The sum rule (\ref{eqn:LFH_CSR}) leads to $U(\zeta_\perp=0) = 0$. 
It was shown that  Eq.~(\ref{eqn:LFSWE}) can be identified with the string equation of motion in the fifth dimension in AdS/QCD if $\zeta_\perp$ is identified with the fifth coordinate $z$ in AdS${}_5$ \cite{Brodsky:2006uqa}. The effective potential $U = (1/4)\Phi'^2 - (1/2)\Phi'' - ({3}/{2z})\Phi'$ is related to the dilaton profile $\Phi(z)$ introduced in AdS/QCD to generate 
confinement by breaking the conformal invariance \cite{Brodsky:2006uqa, Erlich:2005qh}. Phenomenologically, the Regge scaling requires $\Phi(z\to\infty) \to z^2$ \cite{Karch:2006pv}. The chiral condensate $\Sigma=\langle \bar q q\rangle$ is encoded in the background scalar field $X$ at the UV boundary ($z=0$), viz. $\langle X(z\to 0) \rangle \to \frac{1}{2}\Sigma z^3$  in the chiral limit \cite{DaRold:2005mxj}. The former produces a quadratic confining potential, $U(\zeta_\perp\to\infty) \sim \zeta_\perp^2$ in the IR, while the latter requires $U(\zeta_\perp \to 0) \sim -\Sigma^2 \zeta^4_\perp$ in the UV, consistent with the chiral sum rule (\ref{eqn:LFH_CSR}). 

\begin{figure}[h]
\centering
\subfigure[\ wo. $\chi$SB: quadratic potential and LFWF Eq.~(\ref{eqn:sw_pion})]{\includegraphics[width=0.34\textwidth]{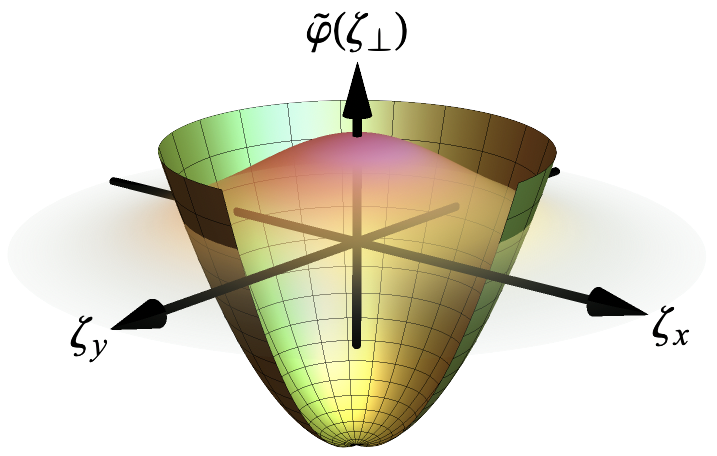}}
\subfigure[\ w. $\chi$SB: sombrero potential and LFWF Eq.~(\ref{eqn:model})]{\includegraphics[width=0.35\textwidth]{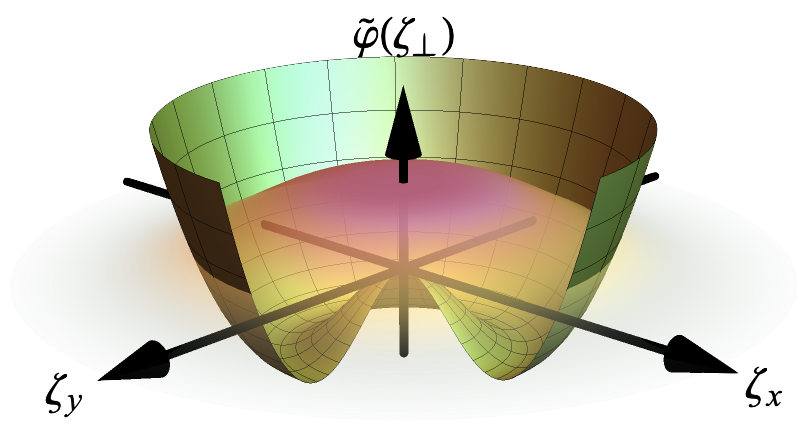}}
\caption{(Colors online) Comparison of the pion wave functions Eq.~(\ref{eqn:sw_pion}) obtained from a purely quadratic potential (\ref{eqn:sw_potential}) without implementing chiral symmetry breaking ($\chi$SB) (\textit{Top panel}) and Eq.~(\ref{eqn:model}) obtained from a sombrero potential (\ref{eqn:pade}) that implements $\chi$SB (\textit{Bottom panel}).
The plateau of the bottom-panel wave function Eq.~(\ref{eqn:model}) near $\zeta_\perp=0$ is required by the chiral sum rule (\ref{eqn:LFH_CSR}). This feature resembles a structureless particle up to the confinement scale.}
\label{fig:transverse}
\end{figure}

The chiral sum rule does not uniquely determine the effective potential. Nevertheless, it can be shown that such potentials have the shape of a sombrero (see Fig.~\ref{fig:transverse}). From the AdS/QCD perspective, this shape is inherited from the Higgs potential through the condensate \cite{Gherghetta:2009ac}. 
The exact shape is dictated by QCD. Instead of 
solving for such a potential, we can construct it from some model pion wave function,  $U = \nabla_\perp^2\widetilde\varphi_\pi/\widetilde\varphi_\pi$. For example, the following harmonic oscillator wave function satisfies the chiral sum rule,
\begin{equation}\label{eqn:model}
\widetilde\varphi_\pi(\zeta_\perp) = \Big[1 + \frac{1}{2}(\kappa\zeta_\perp)^2 + \frac{c}{8}(\kappa\zeta_\perp)^4\Big] e^{-\frac{1}{2}(\kappa\zeta_\perp)^2},
\end{equation}
with a chiral condensate $\Sigma \sim (0.3\,\text{GeV})^3$. 
Here, $\kappa\sim 0.5\,\text{GeV}$ is the confining strength, related to the Regge slope, and $0 < c < 2$ is a dimensionless free parameter unconstrained by the chiral sum rule. Here we chose $c=0.3$ to obtain a good fit to the pion form factor.
The corresponding potential is in Padé form, 
\begin{equation}
U(\zeta_\perp) =\kappa^2\frac{(\kappa\zeta_\perp)^2\big[0.3(\kappa\zeta_\perp)^4+(\kappa\zeta_\perp)^2-11.2\big]}{0.3(\kappa\zeta_\perp)^4+4(\kappa\zeta_\perp)^2+8}. \label{eqn:pade} 
\end{equation}
This is compared with the pion wave function and effective potential from soft-wall AdS/QCD without implementing chiral symmetry breaking \cite{Brodsky:2006uqa, Karch:2006pv},
\begin{align}
& \widetilde\varphi_\pi^\textsc{sw}(\zeta_\perp) = e^{-\frac{1}{2}(\kappa\zeta_\perp)^2}, \label{eqn:sw_pion}\\
& U_\textsc{sw}(\zeta_\perp) = \kappa^4\zeta^2_\perp-2\kappa^2 \label{eqn:sw_potential}
\end{align}
Figure~\ref{fig:transverse} compares the wave function (\ref{eqn:model}) with the Gaussian wave function (\ref{eqn:sw_pion}) obtained from the pure quadratic potential (\ref{eqn:sw_potential}). The former has a distinctive plateau as required by the chiral sum rule. Of course, our model can be improved to accommodate better meson phenomenology \cite{Gherghetta:2009ac, E791:2000xcx}. 

In LFH, the pion valence LFWF [cf.~(\ref{eqn:phi2_LFH})],
\begin{equation}
\psi_\pi(x, \vec k_\perp) = \chi(x)\varphi_\pi(\kappa_\perp).
\end{equation}
Chiral symmetry breaking with the 't~Hooft mechanism suggests a longitudinal function $\chi(x) = \phi_\pi(x)/\sqrt{x(1-x)} \to \text{const.}$ \cite{Li:2021jqb}, resulting in an infinitely long pion \cite{Weller:2021wog}. The valence density, $\widetilde \rho(x, \vec \zeta_\perp) = N |\widetilde\varphi_\pi(\zeta_\perp)|^2$, then also acquires a plateau in the transverse direction ($\vec \zeta_\perp$) due to the chiral sum rule (\ref{eqn:LFH_CSR}). Overall, the partons appear uniformly distributed over the central region of the pion until reaching the confining scale $\zeta_\perp \sim \kappa^{-1}$. 

\begin{figure}[h]
\centering
\includegraphics[width=0.5\textwidth]{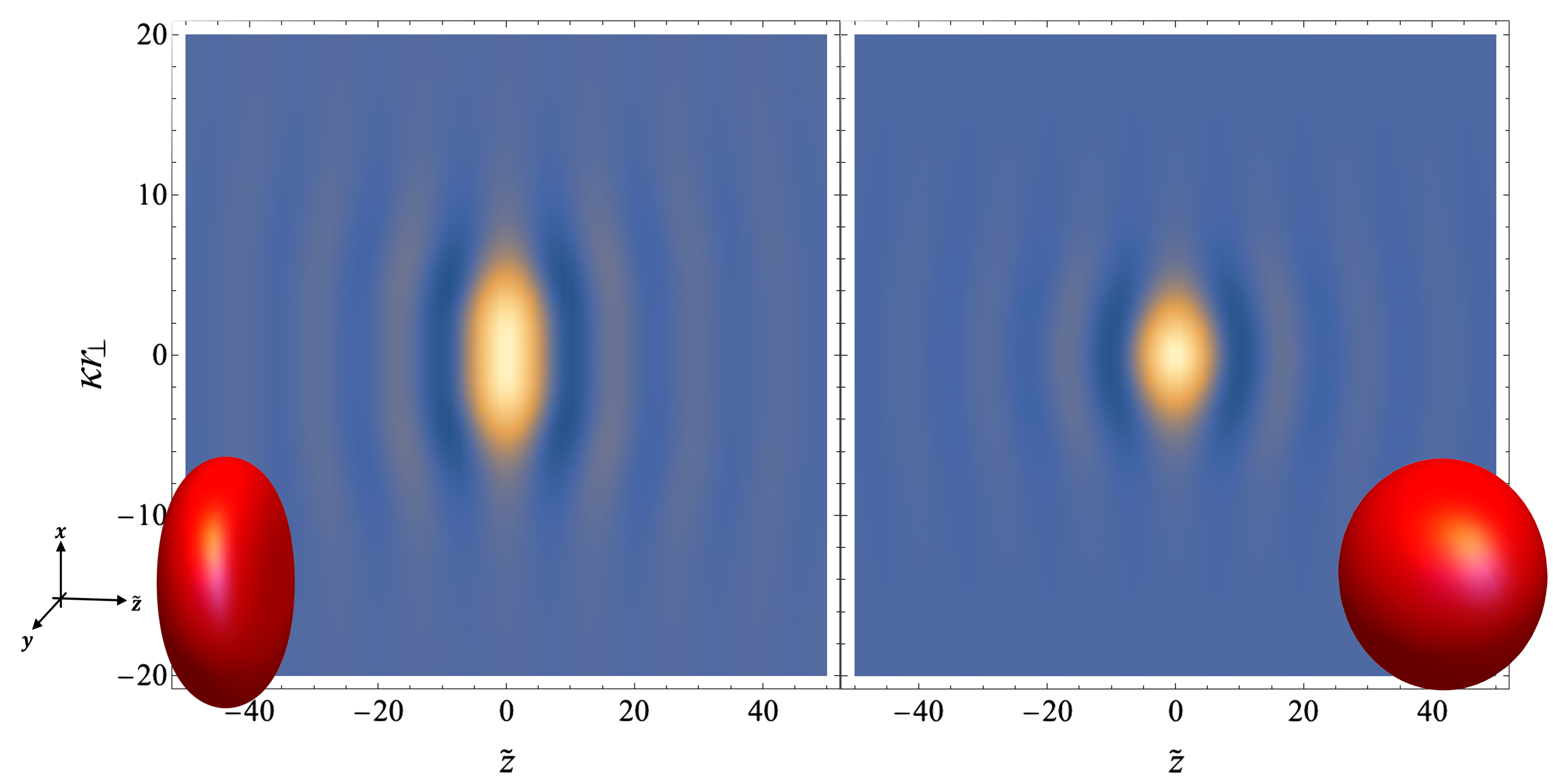}
\caption{(Colors online) Comparison of the 3D coordinate space pion valence wave functions with 
(\textit{Left panel}) and without (\textit{Right panel})  incorporating the chiral symmetry breaking.
These wave functions are obtained from the 3D Fourier transform (\ref{eqn:Ioffe})  of the LFWFs Eqs.~(\ref{eqn:model}) and (\ref{eqn:sw_pion}), respectively. $\vec r_\perp = (x, y)$ are the relative transverse separation of the quark and antiquark, $\tilde z$ is the Ioffe time of Miller and Brodsky. $\kappa$ is the strength of the confining potential.  The pion appears infinitely long in the longitudinal direction ($\tilde z$). The pion with the chiral symmetry breaking (\textit{left}) shows a signature plateau in the transverse direction, also as depicted by the 3D cartoons at the bottom corners.}
\label{fig:separable_model}
\end{figure}

To explore the full 3D structure of the pion, we follow Miller and Brodsky to introduce the coordinate-space pion LFWF as a 3D Fourier transform of the pion LFWF \cite{Miller:2019ysh}, 
\begin{multline}\label{eqn:Ioffe}
\widetilde \psi_\pi(\vec r_\perp, \tilde z) = \int \frac{\dd x}{2\sqrt{x(1-x)}}\int \frac{\dd^2k_\perp}{(2\pi)^3} \\
 \times e^{i(x\tilde z - \vec k_\perp\cdot \vec r_\perp)}
 \psi_\pi(x, \vec k_\perp).
\end{multline}
The third coordinate, $\tilde z = (1/2)p^+_\pi x^-$, is also known as the Ioffe time.
The 3-dimensional pion LFWFs in the coordinate space from these two closely related AdS/QCD models Eqs.~(\ref{eqn:model}, \ref{eqn:sw_pion}) with and without the chiral symmetry breaking, respectively, are compared in Fig.~\ref{fig:separable_model}.  
When the chiral symmetry breaking is taken into account in full 3 dimensions, the pion appears as a uniform disk with long diffractive shadows, a unique impression for subatomic particles yet common for laser beams. 

\begin{figure}[h]
\centering
\includegraphics[width=0.5\textwidth]{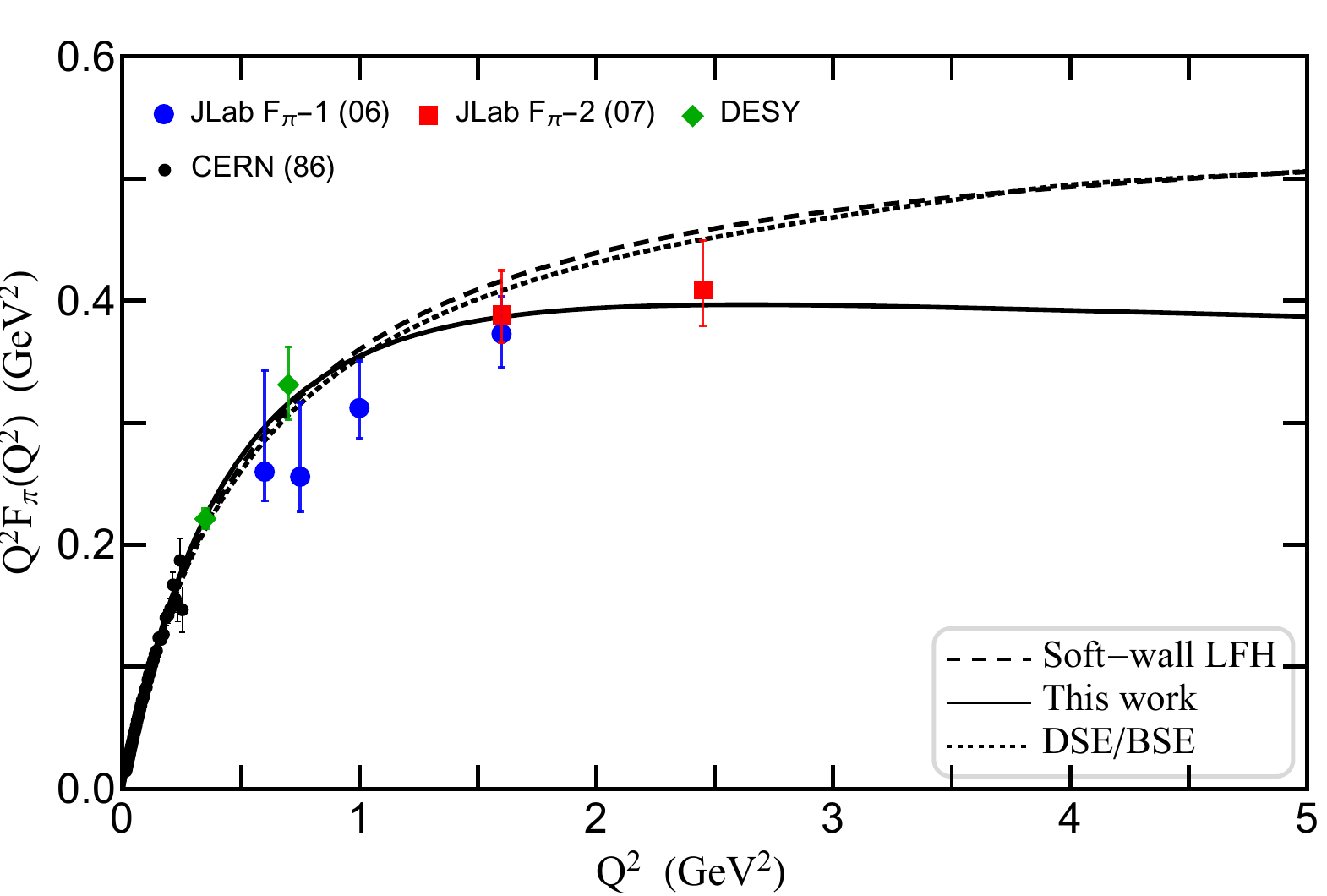}
\caption{(Colors online) The pion electromagnetic form factor obtained using wave function (\ref{eqn:model}) with $\chi$SB and using the soft-wall AdS/QCD wave function (\ref{eqn:sw_pion}) without implementing $\chi$SB. For comparison we also show predictions \cite{Maris:2000sk, Maris:2005tt} from the DSEs, which are frame-independent and consistent with current conservation and $\chi$SB, as well as the experimental measurements \cite{NA7:1986vav, JeffersonLabFpi-2:2006ysh, JeffersonLabFpi:2007vir, JeffersonLab:2008jve} .}
\label{fig:Fpi}
\end{figure}

Figure~\ref{fig:Fpi} compares the pion electromagnetic form factors obtained using 
our wave function (\ref{eqn:model}) and the soft-wall AdS/QCD wave function (\ref{eqn:sw_pion}). We adopt the soft-wall holographic current following Ref.~\cite{Brodsky:2007hb}. The results are compared with the experimental measurements by CERN \cite{NA7:1986vav} and by the Jefferson Lab\cite{JeffersonLabFpi-2:2006ysh, JeffersonLabFpi:2007vir}, including a reanalysis of the DESY data \cite{JeffersonLab:2008jve}. Our results implementing the $\chi$SB improve
the pion form factor at large $Q^2$ as expected.

\section{Summary and discussions \label{sec 5}}

Starting from the axial-vector current conservation and the general structures of the LFWF, we derived a sum rule for the valence wave function of the pion.  This relation implies that the two faces of the pion, a quark-antiquark bound state and a Goldstone boson, are encoded in its inner structure at the large and small parton separation, respectively. 
This real-space image is consistent with the picture from AdS/QCD in five dimensions. This picture is physically reasonable, since the chiral breaking scale $\Lambda_\chi \sim 1\,\text{GeV}$ is larger than the confinement scale $\Lambda_\textsc{qcd} \sim 0.3\text{GeV}$.

Using a concrete analytic model,  we showed that the valence partons are uniformly distributed over the central region of the pion. 
In other words, the pion is approximately a flat disk on the transverse plane with a Gaussian edge. 
Therefore, in this analytic model, the pion is remarkably simple and uniform while encoding both of its desired facets described at the outset. 

The use of the general Lorentz structure of the LFWF is essential to obtain this sum rule. 
In particular, the sum rule constrains the part of wave function associated with the Lorentz structure $\gamma^+\gamma_5/p^+$. Since this part does not appear to be covariant, it is often neglected in the literature (see, e.g., Refs.~\cite{Choi:1997iq, Jaus:1999zv}). In fact, it can be written in a more covariant form: $\slashed{p}\gamma_5$. The emergence of this term signifies the spontaneous chiral symmetry breaking. 
It is also related to the leading-twist pion wave function as well as the associated bi-local matrix element \cite{Burkardt:2002uc}, 
\begin{multline}\label{eqn:Ioffe2}
\langle 0 | \overline q(+\half z) \gamma^+ \gamma_5 q(-\half z)|P(p)\rangle_{z^+=0}  \\
 = 2\sqrt{2N_C}p^+\widetilde \psi_\pi(\vec z_\perp, \tilde z).
\end{multline}
Here, $\tilde z = p\cdot z$ is a generalization of the Ioffe time of Miller and Brodsky \cite{Miller:2019ysh}. 
Only recently, it was realized that this part is important for modeling the pion \cite{Leitner:2010nx, Ahmady:2016ufq, Choi:2020xsr}. 

A proper renormalization is needed for the bare quark mass $m_q$ and the local operators, e.g. $J^\mu_5$. 
The actual implementation of the renormalization is beyond the scope of this work. However, it is important for establishing a full picture of the pion, as well as for practical \textit{ab initio} or model calculations. Thanks to asymptotic freedom, the UV behavior of the pion can be analyzed in perturbation theory, which very useful for the UV renormalization. \cite{Brodsky:1973kr, Lepage:1980fj, Maris:2003vk, Beuf:2016wdz, Lappi:2016oup, Glazek:1994qc}.

The next important issue is the dynamical mass generation. In DSEs, the dynamical quark mass (self-energy) is constrained by the 
AVWTI \cite{Maris:1997hd}. How the AVWTI is realized on the light front is an interesting question. The problem is that the light-front axial charge $Q_5$ does not create the pion pole \cite{Wu:2003vn, Beane:2013oia}.  It is likely that the proper
relation that constrains the dynamical mass generation on the light front comes from additional light-front chiral current algebras \cite{Beane:2013oia}.

There is a long-standing myth that the vacuum in light-front QCD is trivial and hence the pion on the light front is not a Nambu-Goldstone boson. 
This myth has been debunked many times (see, e.g.~Ref.~\cite{Beane:2013oia} for a recent review). The key point is that the chiral condensate on the light front is non-local, viz. $\langle 0 | \bar q q | 0 \rangle = \frac{1}{2}m_q \big\langle 0 \big| \overline q \mathsmaller{{\frac{\gamma^+}{
\tensor\partial^{+}}}} q \big| 0 \big\rangle \ne 0$ in the chiral limit ($m_q \to 0$). However, as Refs.~\cite{Brodsky:2010xf, Brodsky:2022fqy} pointed out, the relevant question is how the chiral condensate can be understood in terms of the pion structure expressed in its light-front wave functions. We hope the Janus-like character of the pion sketched here provides a pathway to an answer.

\section*{Acknowledgements}

Y.L. thanks V.A. Karmanov for enlightening discussions on the covariant structures of the light-front wave functions, and thanks L. Zhang for conversations on the chiral 
symmetry breaking in AdS/QCD. 
This work was supported in part by the US Department of Energy (DOE) under
Grant Nos. DE-FG02-87ER40371, DE-SC0018223 (SciDAC-4/NUCLEI),  and DE-SC0023495 (SciDAC-5/NUCLEI). Y.L. is supported by the New faculty start-up fund of the University of Science and Technology of China. 


\end{document}